\documentstyle[seceq,preprint]{jpsj}

\def\runtitle{
Theory of Tunneling Conductance for Triplet Superconductor Junction}
\def\runauthor{Masashi {\sc Yamashiro} {\it et al.}}

\title
{
Theory of Tunneling Conductance for Normal Metal/Insulator/\\
Triplet Superconductor Junction }

\author
{ 
Masashi {\sc Yamashiro}, Yukio {\sc Tanaka}, Yasunari {\sc Tanuma}
and Satoshi {\sc Kashiwaya}$^{1,}$
}

\inst
{
Graduate School of Science and Technology, 
Niigata University, Ikarashi, Niigata 950-2181\\
$^1$Electrotechnical Laboratory, Tsukuba, Ibaraki 305-9568
}

\recdate
{
\today
}

\abst
{
Tunneling conductance spectra of normal metal/insulator/triplet superconductor
junctions are investigated theoretically. As triplet paring states we select 
several types of symmetries that are promising candidates for the 
superconducting states in UPt$_{3}$ and in Sr$_{2}$RuO$_{4}$. The calculated 
conductance spectra are sensitive to the orientation of the junction which 
reflects the anisotropy of the pairing states. They show either zero-bias 
conductance peaks or gap-like structures depending on the orientation of the 
junctions. The  existence of a residual density of states, peculiar to 
nonunitary states, is shown to have a significant influence on the properties 
of the conductance spectra. Present results serve as a guide for the 
experimental determination of the symmetry of the pair potentials in UPt$_{3}$ 
and Sr$_{2}$RuO$_{4}$.
}

\kword
{
triplet superconductor, nonunitary pair potential, zero-bias 
conductance peak,
}

\begin{document}
\sloppy
\maketitle

\begin{full}
\section{Introduction}
\label{sec:s1}
Since the discovery of superconductivity in
heavy fermion compounds 
the determination of the symmetry of the pair potential in these 
materials 
has been an important issue. \cite{sigrist1}
Among the heavy fermion superconductors, so far
properties of UPt$_{3}$ have been studied most extensively.
Based on NMR experiments\cite{tou} the possibility of 
odd parity pairing states, $i.e.$, triplet pairing states, 
has been suggested. 
Theoretically, several papers propose two-dimensional even parity 
states, 
$i.e.$ singlet pairing states,\cite{park,hess,ueda} 
while others suggest triplet pairing 
states belonging to the one-dimensional representation
(A$_{u}$)\cite{machida1,machida3} 
and to the two-dimensional representation 
(E$_{u}$).\cite{machida4,sauls} 
Although theoretical\cite{fledderjohann,norman,graf} and experimental
\cite{shivaram,ellman,taillfere,lussier,suderow} spectroscopic studies
of the thermal conduction and transverse sound attenuation of UPt$_{3}
$ 
have been performed, the symmetry of the pair potential could not yet 
been
determined. 
\par
Recently superconductivity has been discovered in Sr$_{2}$RuO$_{4}$,
\cite{maeno} which is the first example of 
a noncuprate layered perovskite superconductor. 
Since this compound is isostructual to the cuprate superconductors 
the electronic properties in the normal state\cite{maeno2} 
and superconducting state\cite{yoshida} 
are highly anisotropic. Several experiments\cite{maeno3,ishida} 
indicate a large 
residual density of states of quasiparticles at low temperatures. 
In addition there is evidence for the existence of 
ferromagnetic spin fluctuations. \cite{rice,mazin}
Based on these facts 
two-dimensional triplet superconducting states, belonging to 
the two-dimensional E$_{u}$ symmetry, 
have been proposed for Sr$_{2}$RuO$_{4}$. \cite{sigrist2,machida2} 
Similar to the case of  UPt$_{3}$ 
the symmetry of the pair potential 
of Sr$_{2}$RuO$_{4}$ has not yet been determined.
\par
Phase-sensitive measurements provide the most useful information
for the determination of the symmetry of the pair 
potentials.\cite{Harlingen}
Recently it was shown
that tunneling spectroscopy has the ability to detect the phase of the 
pair potential, \cite{tanaka1,kashiwaya1,kashiwaya2} 
as follows:
In anisotropic superconductors 
quasiparticles feel different signs of the 
pair potentials depending on the directions of their 
motions.\cite{bruder} 
At the normal metal/superconductor interface
the anisotropy of the pair potential significantly influences the 
properties of the Andreev reflections.\cite{andreev,blonder}
As a result tunneling conductance spectra of the 
normal metal/insulator/anisotropic 
superconductor junctions are modified due to the anisotropy of pair 
potential.
\cite{tanaka1,kashiwaya1} 
The most remarkable feature is the existence of 
zero-bias conductance peaks (ZBCP) in the tunneling spectra
for $d$-wave symmetry. 
The origin of these peaks is the 
localized zero energy states (ZES) \cite{hu} 
due to the change of sign of the pair potential in ${\bf k}$-space.
\cite{kashiwaya2} 
ZBCP have actually been observed in
experiments on high-$T_{c}$ 
superconductors\cite{kashiwaya1,geerk,leuser} and 
the consistency between theory and experiments has been checked.
\cite{alff,ueno}
\par
To determine the symmetries of the pair potential of UPt$_{3}$ 
and Sr$_{2}$RuO$_{4}$ it is meaningful to apply the phase sensitive 
capability of tunneling spectroscopy. 
For superconducting UPt$_{3}$
various theoretical and experimental studies of tunneling experiments
have been done,\cite{lohneysen,goll1,nowack,goll2,dewilde,naidyuk} 
but unfortunately effects of the phase have not been considered
in these papers.
Similar to the singlet cases, the appearance of 
ZBCP in conductance spectra
is expected due to the change of sign of the pair potential.
To apply the concepts of phase sensitivity
to triplet superconductors additional studies are required, due to
the large difference in the pairing states.
\par
In the present paper, 
we investigate the tunneling conductance spectra of 
normal metal / insulator/ triplet superconductor 
($N/I/TS$) junctions 
by extending the previous theory 
for anisotropic singlet superconductors. 
Although there are several works concerning surface 
bound states in triplet superconductors \cite{buchholtz,hara}, 
systematic studies for tunnel conductance  in $N/I/TS$ junctions 
have not been performed. 
The following points are of interests:
i) the orientations of ZBCP
which gives essential information about the symmetries,
ii) the spin dependence of the conductance spectra
and iii) the influence of the nonunitary states.
The organization of this paper is as follows: 
In Sec.\ref{sec:s2} a general formula of conductance spectroscopy for 
triplet superconductors is presented. 
Since it is meaningless to present calculations
for all possible triplet symmetries,
the results are given for several typical cases;
promising pairing states for  
UPt$_{3}$ in Sec. \ref{sec:s5} and for Sr$_{2}$RuO$_{4}$ in Sec. 
\ref{sec:s6}. 
Finally in Sec.\ref{sec:s7} we summarize our results and discuss 
future problems.

\section{Conductance formula }
\label{sec:s2}
For the calculation, 
we assume three-dimensional $N/I/TS$ junctions 
with semi-infinite double layer structures 
in the clean limit. 
The flat 
interface is perpendicular to the $x$-axis, and is 
located at $x$=0 [Fig.1(a)].
The barrier potential at the interface 
has a delta-functional form $H\delta(x)$, where $\delta(x)$ 
and $H$ are the delta function and its amplitude, respectively. 
We also consider another orientation of the junction where
the flat interface is perpendicular to the $z$-axis and located at 
$z$=0 [Fig.1(b)].
The Fermi wave number $k_{F}$ and the effective 
mass\cite{lohneysen,deutscher}
{\it m} are assumed to be equal both in the normal metal and 
in the superconductor. 
For simplicity we assume in the following calculations
that the pair potentials are spatially constant. 
The wave function describing the quasiparticles in inhomogeneous 
anisotropic 
superconductors $\Psi(\mbox{\boldmath$r$})$ is 
obtained by the solutions of 
the Bogoliubov-de Gennes (BdG) equation.\cite{bruder,hu} 
After applying the quasi-classical approximation, 
BdG equations are reduced to the  Andreev equation, 
\cite{bruder,hu,kurkijarvi}
\[
\displaystyle
Eu_{s}(\mbox{\boldmath{$r$}})=-iv_{F}\mbox{\boldmath{$k$}}\cdot \nabla 
u_{s}(\mbox{\boldmath{$r$}})+
\sum_{s^{\prime}} \Delta_{ss^{\prime}}(\theta,\phi)
v_{s^{\prime}}(\mbox{\boldmath{$r$}})
\]
\begin{equation}
\displaystyle
\hspace{14pt}
Ev_{s}(\mbox{\boldmath{$r$}})=iv_{F}\mbox{\boldmath{$k$}}\cdot \nabla 
v_{s}(\mbox{\boldmath{$r$}})+\sum_{s^{\prime}} 
\Delta^{\ast}_{ss^{\prime}}(\theta,\phi)
u_{s^{\prime}}(\mbox{\boldmath{$r$}}).
\label{eqn:e21}
\end{equation}
The quantities $u_{s}(\mbox{\boldmath$r$})$ and 
$v_{s}(\mbox{\boldmath$r$})$ are 
electronlike quasiparticles (ELQ) and 
holelike 
quasiparticles (HLQ), respectively with spin index $s$=$\uparrow$ 
or $s$=$\downarrow$. 
In Eq. (\ref{eqn:e21}), \(\mbox{\boldmath{$k$}}\) 
describes the relative motion 
of the Cooper pairs fixed on the Fermi surface 
(\(|\mbox{\boldmath{$k$}}|\) = ${k_{F}}$), 
\( \mbox{\boldmath{$r$}}\) the center of mass coordinates of the pair 
potential and $v_{F}$ the Fermi velocity.
The wave functions $u_{s}(\mbox{\boldmath$r$})$ and 
$v_{s}(\mbox{\boldmath$r$})$ are obtained by neglecting the fast 
oscillating plane-wave part, 
according to the quasiclassical 
approximations.\cite{bruder,kurkijarvi}
In this approximation, we assume that the effective 
pair potential is given by 
\[
{\bf\Delta}(\mbox{\boldmath$k,r$})=
{\bf\Delta}(\theta, \phi)\Theta(x), \hspace{12pt}
\mbox{$z$-$y$ plane interface}\]
\[
\hspace{40pt}={\bf\Delta}(\theta, \phi)\Theta(z), \hspace{12pt}
\mbox{$x$-$y$ plane interface}\]
\begin{equation}
\frac{k_{x}+ik_{y}}{\mid \mbox{\boldmath$k$} 
\mid}=e^{i\phi}\sin\theta, 
\hspace{12pt} \frac{k_{z}}{\mid \mbox{\boldmath$k$} \mid}=\cos\theta
\label{eqn:e22}
\end{equation} 
where $\theta$ is the polar angle 
and $\phi$ is the azimuthal 
angle in the $x$-$y$ plane. 
The quantities $\Theta(x)$ and $\Theta(z)$ 
are the Heaviside step function.
The pair potential matrix is represented  as
\begin{equation}
 {\bf\Delta}(\theta, \phi)=\left( \begin{array}{cc}
\Delta_{\uparrow\uparrow}(\theta, \phi) & \Delta_{\uparrow \downarrow}
(\theta, \phi)\\[10pt]\Delta_{\downarrow \uparrow}(\theta, \phi) & 
\Delta_{\downarrow \downarrow}(\theta, \phi)
\end{array} \right). 
\label{eqn:e23}
\end{equation}
In the present paper, 
we neglect the spin-orbit coupling of the quasiparticles.  
A formula for the tunneling conductance of the $N/I/TS$ 
junction is derived from the reflection and transmission probabilities 
of the electrons injected from the normal metal with angles 
$\theta$ and $\phi$. 
In the following, we assume that 
an  electron is injected with an equal probability  for 
up and down spin. 
The injected electron is reflected as 
an electron (normal reflection) and as a hole (Andreev reflection). 
When the interface is perpendicular to the $x$-axis ($z$-$y$ plane 
interface) 
[Fig.1(a)], the transmitted ELQ and HLQ feel different effective 
pair potentials $\Delta_{ss'}(\theta, \phi_{+})$ and 
$\Delta_{ss'}(\theta, \phi_{-})$, 
with $\phi_{+}=\phi$ and $\phi_{-}=\pi-\phi$. 
On the other hand, in the case when the interface is perpendicular 
to the  $z$-axis ($x$-$y$ plane interface), 
two kinds of quasiparticles feel $\Delta_{ss'}(\theta_{+}, \phi)$ 
and $\Delta_{ss'}(\theta_{-}, \phi)$, 
with $\theta_{+}=\theta$ and $\theta_{-}=\pi-\theta$ [Fig.1(b)], 
respectively. 
The coefficients of the Andreev 
reflection $a_{ss^{\prime}}(E,\theta,\phi)$ and the normal reflection 
$b_{ss^{\prime}}(E,\theta,\phi)$ are determined by solving 
the BdG equation under the following boundary conditions
\begin{equation}
\left. \Psi(\mbox{\boldmath$r$})\right|_{x=0_{-}}=
\left. \Psi(\mbox{\boldmath$r$})\right|_{x=0_{+}},\hspace{20pt}
\displaystyle
\left. \frac{d\Psi(\mbox{\boldmath$r$})}{dx}\right|_{x=0_{-}}=
\left. \frac{d\Psi(\mbox{\boldmath$r$})}{dx}\right|_{x=0_{+}}-
\left. \frac{2mH}{\hbar^{2}}\Psi(\mbox{\boldmath$r$})\right|_{x=0_{-}}
\label{eqn:e24}
\end{equation}
for $z$-$y$ plane interface and 
\begin{equation}
\left. \Psi(\mbox{\boldmath$r$})\right|_{z=0_{-}}=
\left. \Psi(\mbox{\boldmath$r$})\right|_{z=0_{+}},\hspace{20pt}
\displaystyle
\left. \frac{d\Psi(\mbox{\boldmath$r$})}{dz}\right|_{z=0_{-}}=
\left. \frac{d\Psi(\mbox{\boldmath$r$})}{dz}\right|_{z=0_{+}}-
\left. \frac{2mH}{\hbar^{2}}\Psi(\mbox{\boldmath$r$})\right|_{z=0_{-}}
\label{eqn:e25}
\end{equation}
for $x$-$y$ plane interface. 
Using the obtained coefficients, the normalized 
tunneling conductance at zero temperature 
is calculated according to 
the formula given in our previous work\cite{tanaka1,kashiwaya2}
\[
\sigma(E)=
\displaystyle
\frac{\int_0^\pi\!\int_{-\pi/2}^{\pi/2}
[\sigma_{S,\uparrow}(E,\theta,\phi) + 
\sigma_{S,\downarrow}(E,\theta,\phi)]
\sigma_{N}(\theta,\phi)\sin^{2}\theta 
\cos\phi d\theta d\phi}{\int_0^\pi\!\int_{-\pi/2}^{\pi/2}
2\sigma_{N}(\theta,\phi)
\sin^{2}\theta \cos\phi d\theta d\phi}, \hspace{12pt} 
\mbox{$z$-$y$ plane interface}\]
\begin{equation}
\displaystyle
\hspace{38pt}=\frac{\int_0^{\pi/2}\!\int_0^{2\pi}
[\sigma_{S,\uparrow}(E,\theta,\phi) + 
\sigma_{S,\downarrow}(E,\theta,\phi)]
\sigma_{N}(\theta,\phi)\sin\theta \cos\theta d\theta d\phi}
{\int_0^{\pi/2}\!\int_0^{2\pi}
2\sigma_{N}(\theta,\phi)\sin\theta \cos\theta d\theta d\phi}, 
\hspace{12pt} \mbox{$x$-$y$ plane interface}
\label{eqn:e26}
\end{equation}
where $\sigma_{N}(\theta,\phi)$ 
denotes the normal state tunneling conductance given by
\[
\sigma_{N}(\theta,\phi)=
\displaystyle
\frac{\sin^{2}\theta\cos^{2}\phi}{\sin^{2}\theta\cos^{2}\phi+Z^{2}}, 
\hspace{12pt} \mbox{$z$-$y$ plane interface}\]
\[
\displaystyle
\hspace{24pt}=\frac{\cos^{2}\theta}{\cos^{2}\theta+Z^{2}}, 
\hspace{12pt}
\mbox{$x$-$y$ plane interface},\]
\begin{equation}
\hspace{-74pt} Z=\frac{mH}{\hbar^{2}k_{F}}.
\label{eqn:e27}
\end{equation}
Here $Z$ and $E$ denote an effective barrier parameter and 
an energy of quasi-particles measured from the Fermi energy, 
respectively. 
The quantity $\sigma_{S,s}(E,\theta,\phi)$, 
which is the normalized  conductance 
for the specific spin component ($s=\uparrow$ or $\downarrow$) 
for fixed $\theta$ and $\phi$, is given as 
\begin{equation}
\sigma_{S,s}(E,\theta,\phi)=\frac{1+\mid a_{\uparrow s} \mid ^{2} + 
\mid a_{\downarrow s} \mid ^{2}- \mid b_{\uparrow s} \mid ^{2}
- \mid b_{\downarrow s} \mid ^{2}}{\sigma_{N}(\theta,\phi)},
\label{eqn:e28}
\end{equation}
using the 
normal and Andreev reflection coefficients 
$b_{ss^{\prime}}(E,\theta,\phi)$ and $a_{ss^{\prime}}(E,\theta,\phi)
$,
respectively $(s=\uparrow,\downarrow)$. 
Eqs.(\ref{eqn:e26}) and (\ref{eqn:e28}) 
obtained for the conductance are generic. 
Now it is straightforward to obtain 
the tunneling conductance spectra for any triplet superconductor
when the irreducible representations of the 
pair potentials are given. 
For the following discussion, we  point out that 
the low transparency limit of the junction 
means  $\sigma_{N}(\theta,\phi) \rightarrow 0$, 
$i.e.$, $Z \rightarrow \infty$. 

\section{Tunneling spectra in triplet superconductors}
\label{sec:s3}
\subsection{UPt$_{3}$}
\label{sec:s5}
In this section, 
the tunneling conductance of a normal metal / insulator /
UPt$_{3}$ junction is calculated for 
several promising pairing states with 
hexagonal symmetry. 
There are 
one-dimensional\cite{machida1,machida3} (A$_{1u}$, A$_{2u}$) and
two-dimensional\cite{machida4,sauls} (E$_{1u}$, E$_{2u}$) 
representations 
with triplet pairing states and a
two-dimensional E$_{1g}$ state 
\cite{park,hess,ueda} with singlet pairing. 
A brief report of this section was already given 
elsewhere.\cite{yamashiro1}
In the case of a one-dimensional representation, both states, A$_{1u}$ 
and A$_{2u}$, are nonunitary pair potentials. 
\noindent
For both representations the normalized conductance 
$\sigma_{S,\uparrow}(E,\theta,\phi)$ 
is given as 
\begin{equation}
\sigma_{S,\uparrow}(E,\theta,\phi)
\displaystyle
=\frac{1+\sigma_{N}(\theta,\phi)\mid\Gamma\mid^{2}+
[\sigma_{N}(\theta,\phi)-1]\mid\Gamma\mid^{4}}
{\mid1+[\sigma_{N}(\theta,\phi)-1]\Gamma^{2}\mid^{2}}, 
 \hspace{12pt} \mbox{$z$-$y$ plane interface}
\label{eqn:e39}
\end{equation}
\begin{equation}
\displaystyle
\hspace{80pt}=\frac{1+\sigma_{N}(\theta,\phi)\mid\Gamma\mid^{2}+
[\sigma_{N}(\theta,\phi)-1]\mid\Gamma\mid^{4}}
{\mid1-[\sigma_{N}(\theta,\phi)-1]\Gamma^{2}\mid^{2}}, 
 \hspace{12pt} \mbox{$x$-$y$ plane interface}
\label{eqn:e310}
\end{equation}
\[
\displaystyle
\hspace{-36pt}
\Gamma=\frac{E-\Omega}{\mid\Delta_{\uparrow\uparrow}(\theta,\phi)\mid}
,  
\hspace{24pt}\Omega=\sqrt{E^{2}-
\mid\Delta_{\uparrow\uparrow}(\theta,\phi)\mid^{2}}
\]
where $\Delta_{\uparrow\uparrow}(\theta,\phi)$ is given by 

\noindent
A$_{1u}$:
\begin{equation}
\Delta_{\uparrow\uparrow}(\theta,\phi)=\Delta_{0}\cos\theta, \ 
\Delta_{\uparrow\downarrow}(\theta,\phi)=
\Delta_{\downarrow\uparrow}(\theta,\phi)=
\Delta_{\downarrow\downarrow}(\theta,\phi)=0 
\label{eqn:e311}
\end{equation}
\noindent
A$_{2u}$:
\begin{equation}
\Delta_{\uparrow\uparrow}(\theta,\phi)
=\Delta_{0}\cos\theta\sin^{2}\theta, \ 
\Delta_{\uparrow\downarrow}(\theta,\phi)=
\Delta_{\downarrow\uparrow}(\theta,\phi)=
\Delta_{\downarrow\downarrow}(\theta,\phi)=0. 
\label{eqn:e312}
\end{equation}
$\sigma_{S,\downarrow}(E,\theta,\phi)$ 
is equal to unity due to the absence of the effective pair potentials. 
For the $z$-$y$ plane interface junction 
$\sigma(E)$ shows a maximum at $E \sim 0.8\Delta_{0}$ 
and at $E \sim 0.4\Delta_{0}$ 
for the A$_{1u}$ and the A$_{2u}$ states, 
respectively[Fig.\ref{fig:f4}]. 
For the two cases the conductance has a different peak position 
due to the existence of a $\sin^{2} \theta$ factor. 
Reflecting the residual density of states of quasiparticles 
with down spin, 
$\sigma(0)$ converges not to 0, but to 0.5 for the $z$-$y$ plane 
interface 
for $Z \rightarrow \infty$. 
For this case we can show that 
$\sigma(E)$ corresponds to the bulk density 
of states. 
For a $x$-$y$ plane interface junction, 
$\sigma(0)$ increases monotonically with increasing $Z$ 
for both, the A$_{1u}$ and the A$_{2u}$ states. 
In this case,  quasiparticles form ZES on a finite range of the 
Fermi surface (see Table \ref{table2}). \par
The tunneling conductance $\sigma_{S,\downarrow}(E,\theta,\phi)$ of 
the E$_{1u}$ state is given as
\begin{equation}
\sigma_{S,\downarrow}(E,\theta,\phi)
\displaystyle
=\frac{1+\sigma_{N}(\theta,\phi)\mid\Gamma\mid^{2}+
[\sigma_{N}(\theta,\phi)-1]\mid\Gamma\mid^{4}}
{\mid1+e^{-4i\phi}[\sigma_{N}(\theta,\phi)-1]\Gamma^{2}\mid^{2}},
\hspace{12pt} \mbox{$z$-$y$ plane interface}\\[10pt]
\label{eqn:e313}
\end{equation}
\begin{equation}
\displaystyle
\hspace{60pt}=\frac{1+\sigma_{N}(\theta,\phi)\mid\Gamma\mid^{2}+
[\sigma_{N}(\theta,\phi)-1]\mid\Gamma\mid^{4}}
{\mid1-[\sigma_{N}(\theta,\phi)-1]\Gamma^{2}\mid^{2}},
\hspace{12pt} \mbox{$x$-$y$ plane interface}\\[12pt]
\label{eqn:e314}
\end{equation}
\[
\displaystyle
\hspace{-46pt}
\Gamma=\frac{E-\Omega}
{\mid\Delta_{\downarrow\downarrow}(\theta,\phi)\mid},
\hspace{24pt}\Omega=
\sqrt{E^{2}-\mid\Delta_{\downarrow\downarrow}(\theta,\phi)\mid^{2}}.
\]
\begin{equation}
\Delta_{\downarrow\downarrow}(\theta,\phi)
=\Delta_{0}e^{2i\phi}\cos\theta\sin^{2}\theta, \ 
\Delta_{\uparrow\downarrow}(\theta,\phi)=
\Delta_{\downarrow\uparrow}(\theta,\phi)=
\Delta_{\uparrow\uparrow}(\theta,\phi)=0 
\label{eqn:e315}
\end{equation}
\noindent
$\sigma_{S,\uparrow}(E,\theta,\phi)$ 
is equal to unity due to the absence of other pair potentials 
except for $\Delta_{\downarrow\downarrow}(\theta,\phi)$. 
The injected quasiparticles with down spin 
form ZES at the interface depending on the direction of 
their motions 
for $\sigma_{N}(\theta,\phi) \rightarrow 0$. 
For $z$-$y$ plane interface junctions, 
ZES is expected on the line, $\phi=\pm \pi/4$,
whereas for a $x$-$y$ plane interface junction, 
ZES is expected on a finite area of the Fermi surface, 
$0<\theta<\pi/2$[Fig.\ref{fig:f5}]. 

The E$_{2u}$ representation is a unitary state and 
$\sigma(E)$ is given as 
\begin{equation}
\hspace{-20pt}
\sigma_{S,\downarrow}(E,\theta,\phi)=\sigma_{S,\uparrow}(E,\theta,\phi
)
\displaystyle
=\frac{1+\sigma_{N}(\theta,\phi)\mid\Gamma\mid^{2}+
[\sigma_{N}(\theta,\phi)-1]\mid\Gamma\mid^{4}}
{\mid1+e^{-4i\phi}[\sigma_{N}(\theta,\phi)-1]\Gamma^{2}\mid^{2}}, 
 \hspace{12pt} \mbox{$z$-$y$ plane interface}
\label{eqn:e316}
\end{equation}
\begin{equation}
\displaystyle
\hspace{102pt}=\frac{1+\sigma_{N}(\theta,\phi)\mid\Gamma\mid^{2}+
[\sigma_{N}(\theta,\phi)-1]\mid\Gamma\mid^{4}}
{\mid1-[\sigma_{N}(\theta,\phi)-1]\Gamma^{2}\mid^{2}}, 
 \hspace{12pt} \mbox{$x$-$y$ plane interface}\\[12pt]
\label{eqn:e317}
\end{equation}
\[
\displaystyle
\Gamma=\frac{E-\Omega}{\mid\Delta_{\uparrow\downarrow}(\theta,\phi)
\mid}, 
\hspace{24pt}\Omega=\sqrt{E^{2}-
\mid\Delta_{\uparrow\downarrow}(\theta,\phi)\mid^{2}}.
\]
\begin{equation}
\Delta_{\uparrow\downarrow}(\theta,\phi)=
\Delta_{\downarrow\uparrow}(\theta,\phi)=
\Delta_{0}e^{2i\phi}\cos\theta\sin^{2}\theta , 
\hspace{12pt}
\Delta_{\uparrow\uparrow}(\theta,\phi)=
\Delta_{\downarrow\downarrow}(\theta,\phi)=
0.
\label{eqn:e318}
\end{equation}
The line shape of $\sigma(E)$ is similar to that of the
E$_{1u}$ state 
since both pair potentials have the same orbital dependence. 
However, $\sigma(E)$ for the E$_{1u}$ state is 
always larger than 0.5 due to the nonunitarity[Fig.\ref{fig:f5}]. \par
Finally,  let us consider another promising state, 
$i.e.$, the E$_{1g}$ state, where the pair potential is a singlet. 
Applying the previous theory by Tanaka and Kashiwaya 
\cite{tanaka1,kashiwaya1,kashiwaya2,tanaka2}, 
$\sigma_{S,\uparrow}(E,\theta,\phi)$ and 
$\sigma_{S,\downarrow}(E,\theta,\phi)$ are obtained as 
\begin{equation}
\hspace{-20pt}
\sigma_{S,\uparrow}(E,\theta,\phi)=\sigma_{S,\downarrow}(E,\theta,\phi
)
\displaystyle
=\frac{1+\sigma_{N}(\theta,\phi)\mid\Gamma\mid^{2}+
[\sigma_{N}(\theta,\phi)-1]\mid\Gamma\mid^{4}}
{\mid1-e^{-2i\phi}[\sigma_{N}(\theta,\phi)-1]\Gamma^{2}\mid^{2}}, 
 \hspace{12pt} \mbox{$z$-$y$ plane interface}
\label{eqn:e319}
\end{equation}
\begin{equation}
\displaystyle
\hspace{102pt}=\frac{1+\sigma_{N}(\theta,\phi)\mid\Gamma\mid^{2}+
[\sigma_{N}(\theta,\phi)-1]\mid\Gamma\mid^{4}}
{\mid1-[\sigma_{N}(\theta,\phi)-1]\Gamma^{2}\mid^{2}}, 
 \hspace{12pt} \mbox{$x$-$y$ plane interface}\\[12pt]
\label{eqn:e320}
\end{equation}
\[
\displaystyle
\Gamma=\frac{E-\Omega}{\mid\Delta(\theta,\phi)\mid},  
\hspace{24pt}\Omega=\sqrt{E^{2}-\mid\Delta(\theta,\phi)\mid^{2}}, 
\]
with $\Delta(\theta,\phi)=\Delta_{0}e^{2i\phi}\cos\theta\sin\theta$. 
For $z$-$y$ plane interface junctions, 
ZES is expected on the line, $\phi=0$,
whereas
for a $x$-$y$ plane interface junction, 
ZES is expected on a finite area of the Fermi surface, 
$0<\theta<\pi/2$. 
[Fig.\ref{fig:f6}]. \par
All pair potentials  discussed in 
this section are promising candidates 
for the superconducting state in UPt$_{3}$. 
For all of them ZBCP appear when the interface 
is perpendicular to the $z$-axis. Performing tunneling 
spectroscopy measurements on a high purity UPt$_{3}$ sample with 
a well oriented junction, 
ZBCP are expected to be  observed for the junction with a
$x$-$y$ plane interface. 
We strongly hope that ZBCP will be observed in the actual experiments 
in near future. 
For the case of a $z$-$y$ plane interface, 
a wide variety of line shapes of the 
conductance are expected, because of
the irreducible representations of the pair potentials and the 
direction of the junctions.

\subsection{Sr$_{2}$RuO$_{4}$}
\label{sec:s6}
\noindent
As a first case of this section, 
we consider two nonunitary pair potentials. 
Previously, we have already reported the tunneling spectroscopy 
in superconducting Sr$_{2}$RuO$_{4}$.\cite{yamashiro2} 
Since this material has two-dimensional electronic properties, 
we used a nearly cylindrical Fermi surface. 
The resultant conductance formula (\ref{eqn:e26}) is reduced to 
Eqs.(4) to (6) in Ref. 51. In this previous paper we chose two kinds 
of nonunitary pair potentials with tetragonal symmetry, 
$\Delta_{\uparrow\uparrow}(\theta,\phi)=
\Delta_{0}\sin\theta(\sin\phi+\cos\phi)$ (E$_{u}$(1) state) and 
$\Delta_{\uparrow\uparrow}(\theta,\phi)=\Delta_{0}e^{i\phi}\sin\theta$ 
(E$_{u}$(2) state). 
The obtained conductance formulas were given by Eqs.(7) to (10) and 
spectra were shown in Figs.2 and 3 in Ref. 51. 
In a $x$-$y$ plane interface junction, 
$\sigma(E)$ coincides with the bulk density of states 
for both, the E$_{u}(1)$ and the E$_{u}$(2) states.
In the E$_{u}$(2) case, 
U-shaped spectra are obtained [Fig.3 in Ref. 51] 
due to the nearly two-dimensionality of the system. 
In the case of a $z$-$y$ plane interface junction 
with low transparency, 
ZES is expected for $-\pi/4 \leq \phi \leq \pi/4$ for the  
E$_{u}$(1) state, 
while in the case of the E$_{u}$(2) state, ZES is expected for
$\phi=0$. 
Consequently, $\sigma(0)$ of E$_{u}$(1) diverges with increasing
$Z$, while $\sigma(0)$ of the E$_{u}$(2) converges to 
a finite value. 
Our results are consistent with recent results 
by Honerkamp $et$ $al.$\cite{honerkamp} 
As a second case, 
we calculate tunneling conductance spectra for Sr$_{2}$RuO$_{4}$ with 
unitary pair potentials which are given as 
$\Delta_{\uparrow\downarrow}=\Delta_{0}\sin\theta(\sin\phi+\cos\phi)$ 
(E$_{u}$(U1) state) and 
$\Delta_{\uparrow\downarrow}=\Delta_{0}e^{i\phi}\sin\theta$
(E$_{u}$(U2) state). For both states, 
$\Delta_{\uparrow\uparrow}$ and $\Delta_{\downarrow\downarrow}=0$ and
$\Delta_{\uparrow\downarrow}=\Delta_{\downarrow\uparrow}$ are 
satisfied. 
A conductance formulas for these pair potentials are given as 

\vspace{12pt}

\noindent
E$_{u}$(U1):
\[
\hspace{-11.5cm}
\sigma_{S,\uparrow}(E,\theta,\phi)=
\sigma_{S,\downarrow}(E,\theta,\phi)
\]
\begin{equation}
\displaystyle
=\frac{1+\sigma_{N}(\theta,\phi)\mid\Gamma_{+}\mid^{2}+
[\sigma_{N}(\theta,\phi)-1]\mid\Gamma_{+}\mid^{2}\mid\Gamma_{-}\mid^{2
}}
{\mid1+[\sigma_{N}(\theta,\phi)-1]\Gamma_{+}\Gamma_{-}\mid^{2}}, 
 \hspace{12pt} \mbox{$z$-$y$ plane interface}
\label{uppe322}
\end{equation}
\begin{equation}
\displaystyle
=\frac{1+\sigma_{N}(\theta,\phi)\mid\Gamma_{+}\mid^{2}+
[\sigma_{N}(\theta,\phi)-1]\mid\Gamma_{+}\mid^{4}}
{\mid1+[\sigma_{N}(\theta,\phi)-1]\Gamma_{+}^{2}\mid^{2}}, 
 \hspace{12pt} \mbox{$x$-$y$ plane interface}\\[12pt]
\label{uppe323}
\end{equation}
\[
\displaystyle
\Gamma_{\pm}=\frac{\Delta_{0}\sin\theta(\sin\phi\pm\cos\phi)}
{E+\Omega_{\pm}},  
\hspace{24pt}\Omega_{\pm}=
\sqrt{E^{2}-\mid\Delta_{0}\sin\theta(\sin\phi\pm\cos\phi)\mid^{2}}.
\]

\vspace{24pt}

\noindent
E$_{u}$(U2):
\[
\hspace{-10.8cm}
\sigma_{S,\uparrow}(E,\theta,\phi)=\sigma_{S,\downarrow}(E,\theta,\phi
)
\]
\begin{equation}
\displaystyle
=\frac{1+\sigma_{N}(\theta,\phi)\mid\Gamma\mid^{2}+
[\sigma_{N}(\theta,\phi)-1]\mid\Gamma\mid^{4}}
{\mid1-e^{-2i\phi}[\sigma_{N}(\theta,\phi)-1]\Gamma^{2}\mid^{2}}, 
 \hspace{12pt} \mbox{$z$-$y$ plane interface}
\label{uppe324}
\end{equation}
\begin{equation}
\displaystyle
=\frac{1+\sigma_{N}(\theta,\phi)\mid\Gamma\mid^{2}+
[\sigma_{N}(\theta,\phi)-1]\mid\Gamma\mid^{4}}
{\mid1+[\sigma_{N}(\theta,\phi)-1]\Gamma^{2}\mid^{2}}, 
 \hspace{12pt} \mbox{$x$-$y$ plane interface}\\[12pt]
\label{uppe325}
\end{equation}
\[
\displaystyle
\hspace{-50pt}
\Gamma=\frac{E-\Omega}
{\mid\Delta_{\uparrow\uparrow}(\theta,\phi)\mid}, 
\hspace{24pt}\Omega=\sqrt{E^{2}-
\mid\Delta_{\uparrow\uparrow}(\theta,\phi)\mid^{2}}.
\]
The obtained spectra are similar to those of 
the E$_{u}$(1) and E$_{u}$(2) cases 
as shown in Fig.\ref{fig:f8}. 
However, $\sigma(0)$ vanishes with the increasing $Z$ 
in the case of $x$-$y$ plane interface due to the absence of 
residual density of states of quasiparticles.\par
Throughout this paper the $\delta$-function model 
has been used to express the insulating barrier. 
For the junctions of singlet superconductors 
we have already confirmed that 
if the finite thickness of the insulating 
barrier is taken into account 
the essential results of the $\delta$-function model, 
$e.g.$, the conditions of the  existence of the zero energy states, 
are not changed at all.\cite{tanaka1,tanaka3} 
For the junctions of triplet superconductors 
it is interesting to calculate the tunneling conductance in a model
where the thickness of the insulating region is given as $d_{i}$. 
Then only the tunneling conductance in the normal state 
is modified to 
\begin{equation}
\sigma_{N}(\theta,\phi)=
\displaystyle
\frac{4Z^{2}(\theta,\phi)}
{[1-Z^{2}(\theta,\phi)]^{2}\sinh^{2}(\lambda d_{i})+
4Z^{2}(\theta,\phi)\cosh^{2}(\lambda d_{i})},
\label{eqn:e41}
\end{equation}
\[ \lambda=\left[1-\kappa^{2}\alpha^{2}(\theta,\phi)\right]^{\frac12}
\lambda_{0},\hspace{12pt}
Z(\theta,\phi)=\frac{\kappa\alpha(\theta,\phi)}
{\sqrt{1-\kappa^{2}\alpha^{2}(\theta,\phi)}}\]
\begin{equation}
\lambda_{0}=\sqrt{\frac{2mU_{0}}{\hbar}},\hspace{12pt}
\kappa=\frac{k_{F}}{\lambda_{0}}.
\label{eqn:e42}
\end{equation}
In the above, $U_{0}$ is the magnitude of the Hartree potential 
in the insulating region. 
The form factors $\alpha(\theta,\phi)$ are given as 
$\alpha(\theta,\phi)=\sin\theta\cos\phi$ for the $z$-$y$ plane 
interface and 
$\alpha(\theta,\phi)=\cos\theta$ for the $x$-$y$ plane interface. 
The resulting tunneling spectra for this normal conductance are shown 
in Fig.\ref{fig:f7} for the examples of E$_{u}$(1) and E$_{u}$(2). 
The obtained line shapes of the tunneling conductance 
are almost similar to those in  Figs.2 and 3 in Ref. 51. 
The increase of the thickness of the insulating barrier 
corresponds to that of $Z$ in the $\delta$-function model.

\section{Conclusions}
\label{sec:s7}
The tunneling conductance spectra in normal metal/insulator/triplet
superconductor junctions have been presented for various promising 
pairing 
states for the superconductors 
UPt$_{3}$ and Sr$_{2}$RuO$_{4}$. 
The obtained conductance spectra 
exhibit very fruitful features, which 
in the low transparency limit can be classified into six cases: 
\par
(1) The first case is that  $\sigma(0)$ monotonically
increases with increasing $Z$. 
ZES is formed in a finite area on the Fermi surface. 
(E$_{2u}$ state with a $x$-$y$ plane interface, 
E$_{1g}$ state with a $x$-$y$ plane interface, 
E$_{u}$(U1) state with a $z$-$y$ plane interface)  \par
(2) The second case is that $\sigma(0)$ converges to zero for 
sufficiently large $Z$. 
ZES does not appear on the Fermi surface.
(E$_{u}$(U1) state with a $x$-$y$ plane interface, 
E$_{u}$(U2) state with a $x$-$y$ plane interface) \par
(3) The third case is that $\sigma(0)$ converges to 
nonzero value with increasing $Z$. 
Here ZES is formed on a certain line on the Fermi surface. 
(E$_{2u}$ state with a $z$-$y$ plane interface, 
E$_{1g}$ state with a $z$-$y$ plane interface, 
E$_{u}$(U2) state with a $z$-$y$ plane interface) \par
(4) The fourth case are nonunitary states 
with ZBCP. 
(A$_{1u}$ state with a $x$-$y$ plane interface, 
A$_{2u}$ state with a $x$-$y$ plane interface, 
E$_{1u}$ state with a $x$-$y$ plane interface, 
E$_{u}$(1) state with a $z$-$y$ plane interface) \par
(5) The fifth case are nonunitary states where 
for sufficiently larger $Z$, 
$\sigma(0)$ converges to 0.5. 
(A$_{1u}$ state with a $z$-$y$ plane interface, 
A$_{2u}$ state with a $z$-$y$ plane interface, 
E$_{u}$(1) state with a $x$-$y$ plane interface, 
E$_{u}$(2) state with a $x$-$y$ plane interface)  \par
(6) The sixth case are nonunitary states where 
$\sigma(E)$ converges to a certain finite value 
larger than 0.5. 
(E$_{1u}$ state a with $z$-$y$ plane interface, E$_{u}$(2) state with 
a $z$-$y$ plane interface) \par
\noindent
The regions of the Fermi surface which contribute to the 
ZES is summarized in Tables \ref{table2} to \ref{table3}. 
Our classification can serve as a guide in the determination of 
the symmetry of the pair potentials in UPt$_{3}$ and 
Sr$_{2}$RuO$_{4}$. 
We hope that tunneling spectroscopy experiments 
will be performed on a very clean sample 
using a well oriented junction with flat interfaces. 
In the determination of the symmetry of the unknown pair potentials
the barrier height and the directions of the tunneling conductance 
are the most important points. \par
Finally, we comment on the gauge invariance concerning 
the condition of ZES. For this purpose, 
let us discuss the role of 
external phase $\chi_{0}$ and internal phase $\varphi^{i}$ 
in anisotropic superconductors. 
In general, pair potential 
$\Delta_{ss'}(\theta,\phi)$ can be expressed as 
$\Delta_{ss'}(\theta,\phi) 
= \mid \Delta_{ss'}(\theta,\phi) \mid 
\exp[i(\varphi^{i} + \chi_{0})]$. 
The internal phase $\varphi^{i}$ 
of the pair potential is measured from the $c$- and 
$a$-axes using polar and azimuthal angle, 
$\theta$ and $\phi$ throughout this paper. 
The quantity $\varphi^{i}$ depends on 
the direction of the motion of the quasiparticles. 
On the other hand, global phase $\chi_{0}$ is 
independent of 
the direction of the motion of the quasiparticles. \par
To discuss the physical meaning of these two phases 
more clearly, let us consider a trajectory of a 
quasiparticle injected from the superconductor to 
the interface. 
The quasiparticle does not always 
feel  the same phase of the pair potential 
before and after the reflection 
\cite{tanaka1,kashiwaya1,kashiwaya2}. 
We denote 
the phase felt by the quasiparticle before and after the reflection 
as $\bar{\varphi}_{1}$ and $\bar{\varphi}_{2}$, respectively, 
with $\bar{\varphi}_{1}=\varphi^{i}_{1} + \chi_{0}$
and $\bar{\varphi}_{2}=\varphi^{i}_{2} + \chi_{0}$, 
where $\varphi^{i}_{1}$ and $\varphi^{i}_{2}$ are 
the internal phases of the pair potentials. 
The condition of the ZES is determined by the 
phase increment felt by the quasiparticle, 
$\bar{\varphi}_{1} -\bar{\varphi}_{2}$. 
It should be remarked that 
this condition is determined by 
$\varphi^{i}_{1} -\varphi^{i}_{2}$ 
and is regardless of the choice of $\chi_{0}$. 
Consequently, we can verify the  
U(1)-gauge invariance of the condition of ZES. \par
In this paper, the effects of the spatial dependence of the pair 
potential 
have not been determined self-consistently.  
In anisotropic superconductors, the spatial dependence of the pair 
potentials 
is important for a quantitative discussion.\cite{barash} 
For some cases a coexistence of pair potentials is expected.
\cite{matsumoto}
A coexistence of several pair potentials at the interface 
is expected to influence the tunneling conductance. 
Furthermore, the influence of the roughness of the surface 
on the conductance is also an interesting problem.\cite{yamada,nagato,
tanuma} 

\section*{Acknowledgments}
We would like to thank Prof. A. Doenni for critical reading of our 
paper. 
One of the authors (Y.T.) is supported by a 
Grant-in-Aid for Scientific Research in Priority Areas, 
"Anomalous metallic states near the Mott transition," 
and "Nissan Science Foundation".
The computation in this work has been done using the facilities of the 
Supercomputer Center, Institute for Solid State Physics, 
University of Tokyo.

\newpage

\begin{fulltable}
\caption{The region on the Fermi surface contributing to ZES 
for UPt$_{3}$
\label{table2}}
\begin{fulltabular}{@{\hspace{\tabcolsep}\extracolsep{\fill}}ccc} 
\hline
& $z$-$y$ plane interface & $x$-$y$ plane interface\\ \hline
A$_{1u}$,\hspace{2pt}A$_{2u}$ & none & 
$0\leq\phi\leq2\pi$,\hspace{4pt}$0\leq\theta\leq\pi/2$ \\
E$_{1u}$,\hspace{2pt}E$_{2u}$ & $\phi=\pm\pi/4$ &
$0\leq\phi\leq2\pi$,\hspace{4pt}$0\leq\theta\leq\pi/2$ \\
E$_{1g}$ & $\phi=0$ & 
$0\leq\phi\leq2\pi$,\hspace{4pt}$0\leq\theta\leq\pi/2$ \\ \hline
\end{fulltabular}
\end{fulltable}


\begin{fulltable}
\caption{The region on the Fermi surface contributing to ZES 
for Sr$_{2}$RuO$_{4}$
\label{table3}}
\begin{fulltabular}{@{\hspace{\tabcolsep}\extracolsep{\fill}}ccc} 
\hline
&$z$-$y$ plane interface&$x$-$y$ plane interface\\ \hline
E$_{u}$(1),\hspace{2pt}E$_{u}$(U1)&$-\pi/4\leq\phi\leq\pi/4$&none\\
E$_{u}$(2),\hspace{2pt}E$_{u}$(U2)&$\phi=0$&none\\ \hline
\end{fulltabular}
\end{fulltable}


\begin{fullfigure}
\caption{(a): Schematic illustration of the reflection and the 
transmission 
process of the quasiparticle at the interface of the 
junction with $z$-$y$ plane interface 
and (b): $x$-$y$ plane interface. 
The $\theta$ and $\phi$ are the polar angle and azimuthal angle, 
respectively.}
\label{fig:f0}
\end{fullfigure}

\begin{fullfigure}
\caption{Normalized conductance for (a) the A$_{1u}$ state and (b) the
A$_{2u}$ state 
when the interface is perpendicular to the $x$- and the $z$-axes, 
as indicated in the figure. 
Effective barrier parameters are a:$Z$=0.1, b:$Z$=1 and c:$Z$=5.}
\label{fig:f4}
\end{fullfigure}

\begin{fullfigure}
\caption{Normalized conductance for (a) the E$_{1u}$ state and (b) the
E$_{2u}$ state 
when the interface is perpendicular to the $x$- and the $z$-axes, 
as indicated in the figure. 
Effective barrier parameters are a:$Z$=0.1, b:$Z$=1 and c:$Z$=5.}
\label{fig:f5}
\end{fullfigure}

\begin{fullfigure}
\caption{Normalized conductance for the E$_{1g}$ state 
when the interface is perpendicular to the $x$- and the $z$-axes, 
as indicated in the figure. 
Effective barrier parameters are a:$Z$=0.1, b:$Z$=1 and c:$Z$=5.}
\label{fig:f6}
\end{fullfigure}

\begin{fullfigure}
\caption{Normalized conductance for (a) the E$_{u}$(U1) state and 
(b) the E$_{u}$(U2) state 
when the interface is perpendicular to the $x$- and the $z$-axes, 
as indicated in the figure. 
Effective barrier ($\delta$-function model)
parameters are a:$Z$=0.1, b:$Z$=1 and c:$Z$=5.}
\label{fig:f8}
\end{fullfigure}

\begin{fullfigure}
\caption{Normalized conductance for (a) the E$_{u}$(1) state and 
(b) the E$_{u}$(2) state when the insulator has a finite thickness 
$d_{i}$ for $\kappa=0.5$ with 
a:$\lambda_{0}d_{i}$=0.1, b:$\lambda_{0}d_{i}$=1 and 
c:$\lambda_{0}d_{i}$=5.}
\label{fig:f7}
\end{fullfigure}

\end{full}
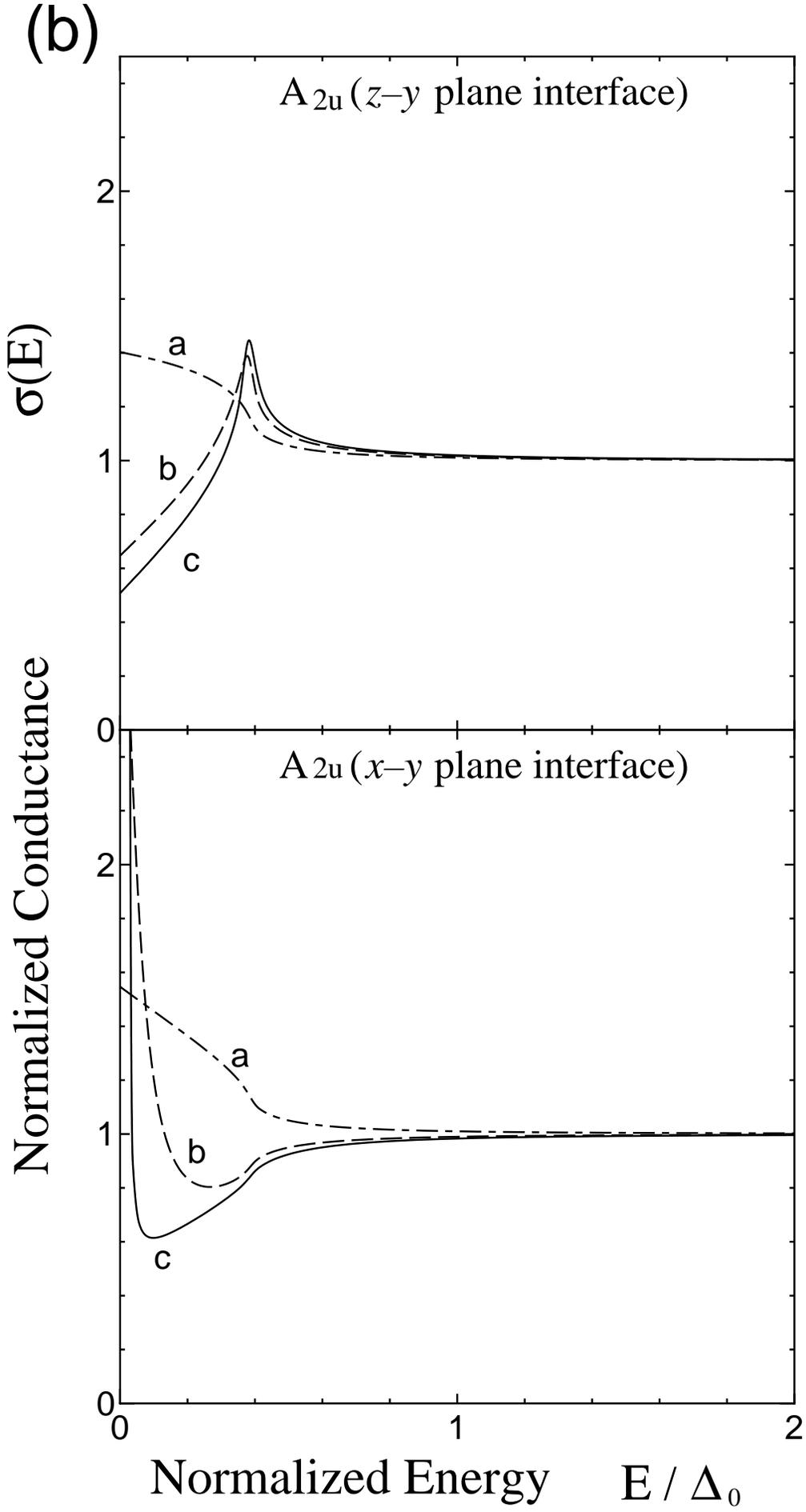
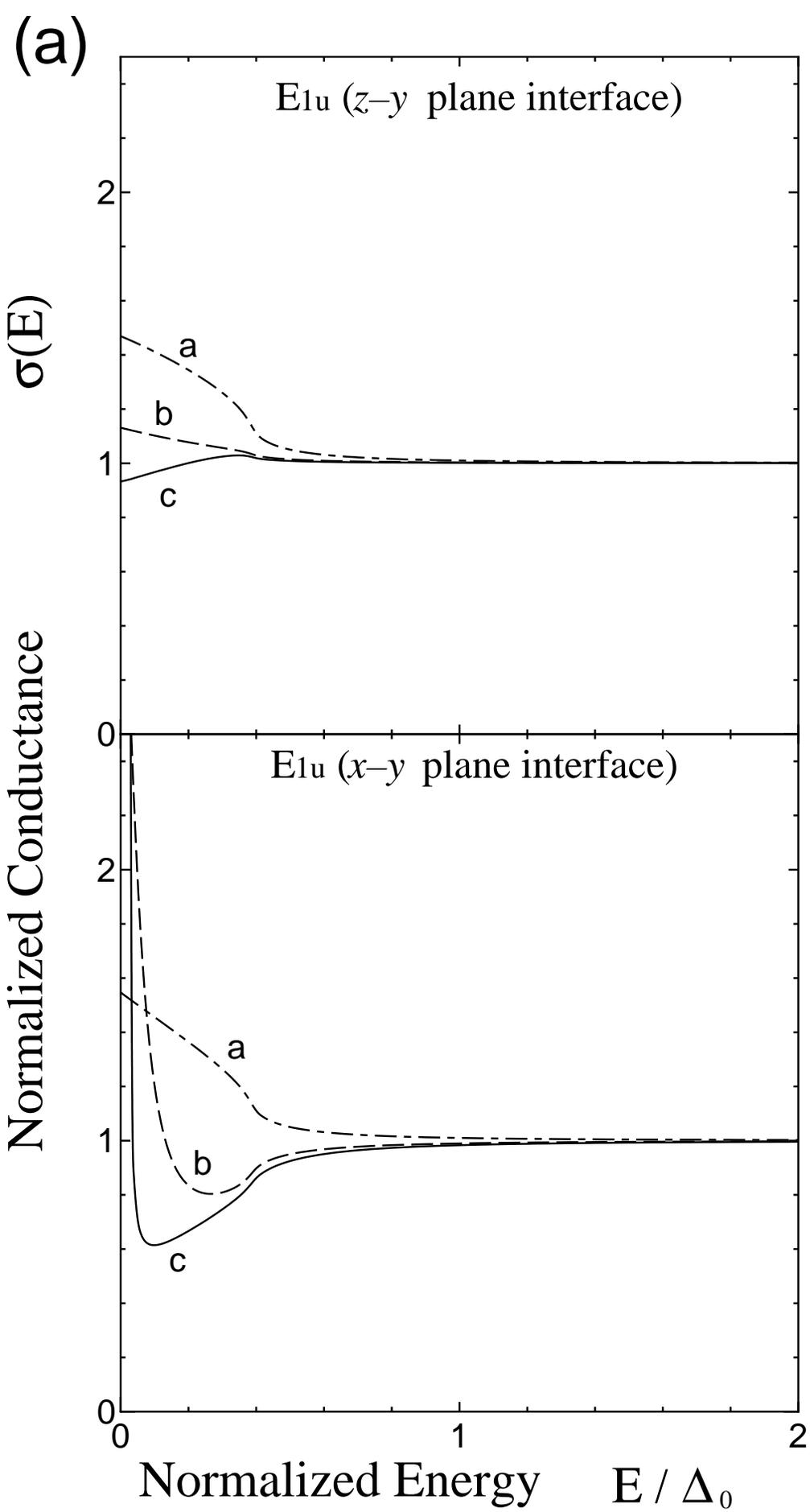
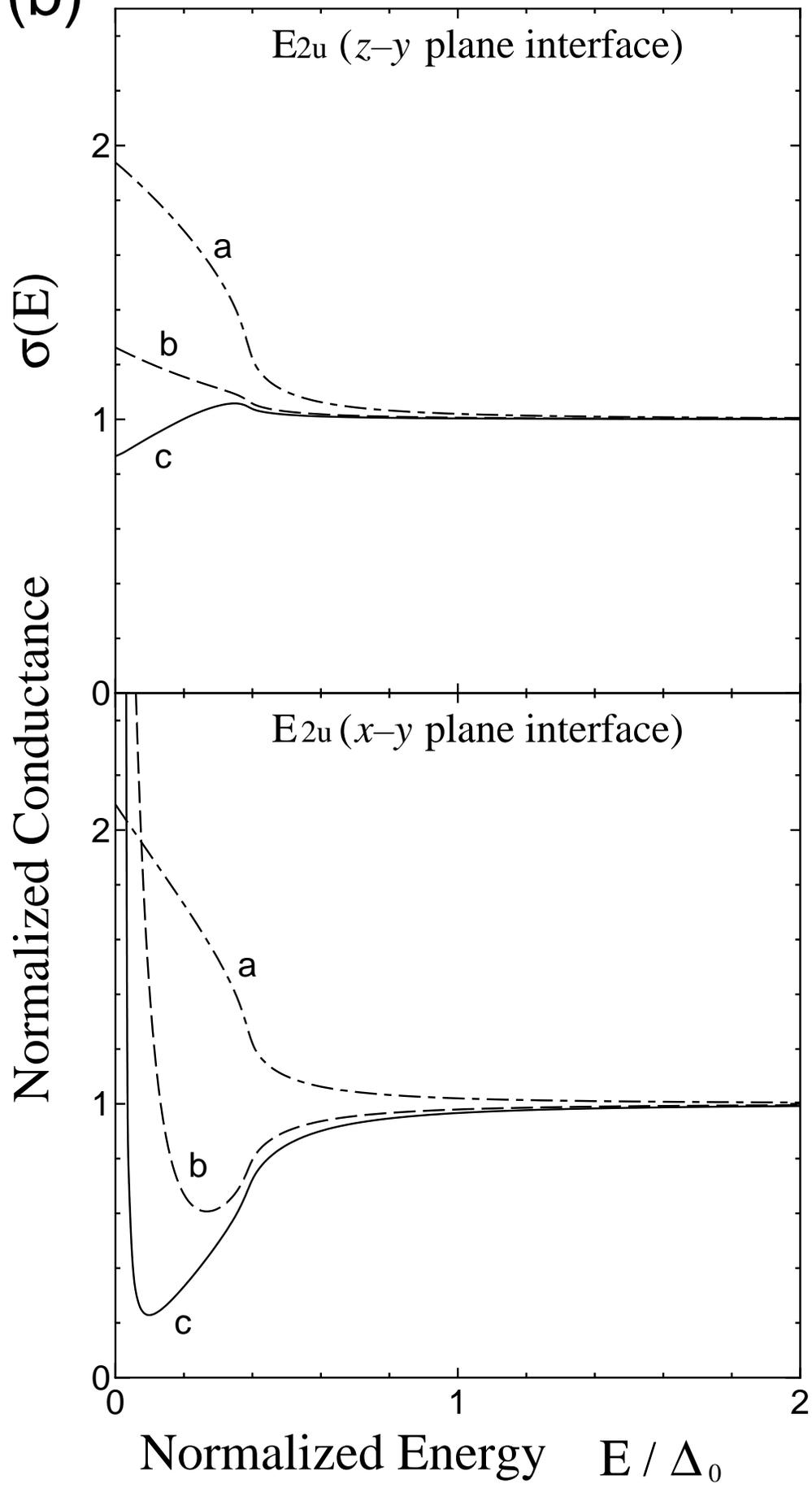
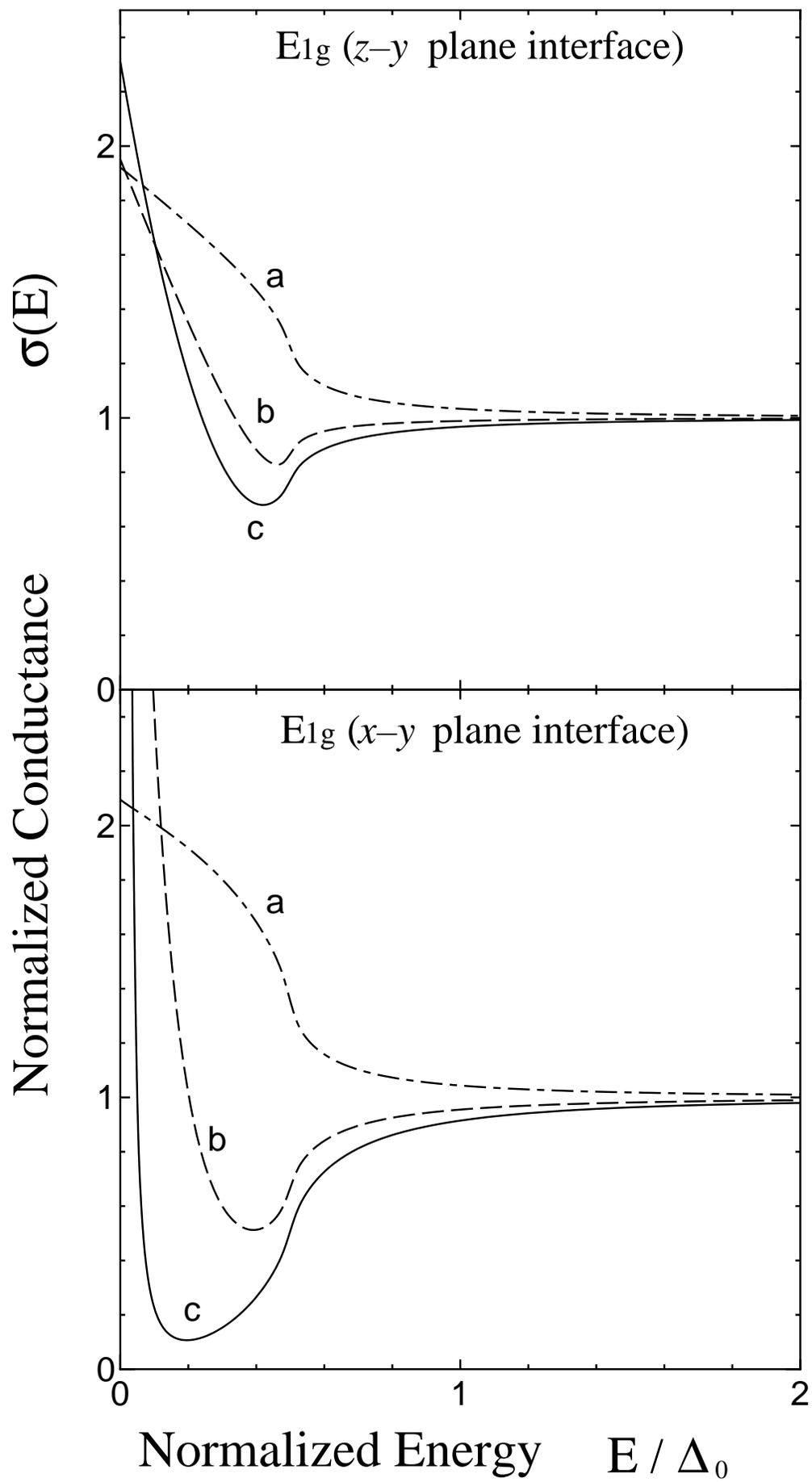
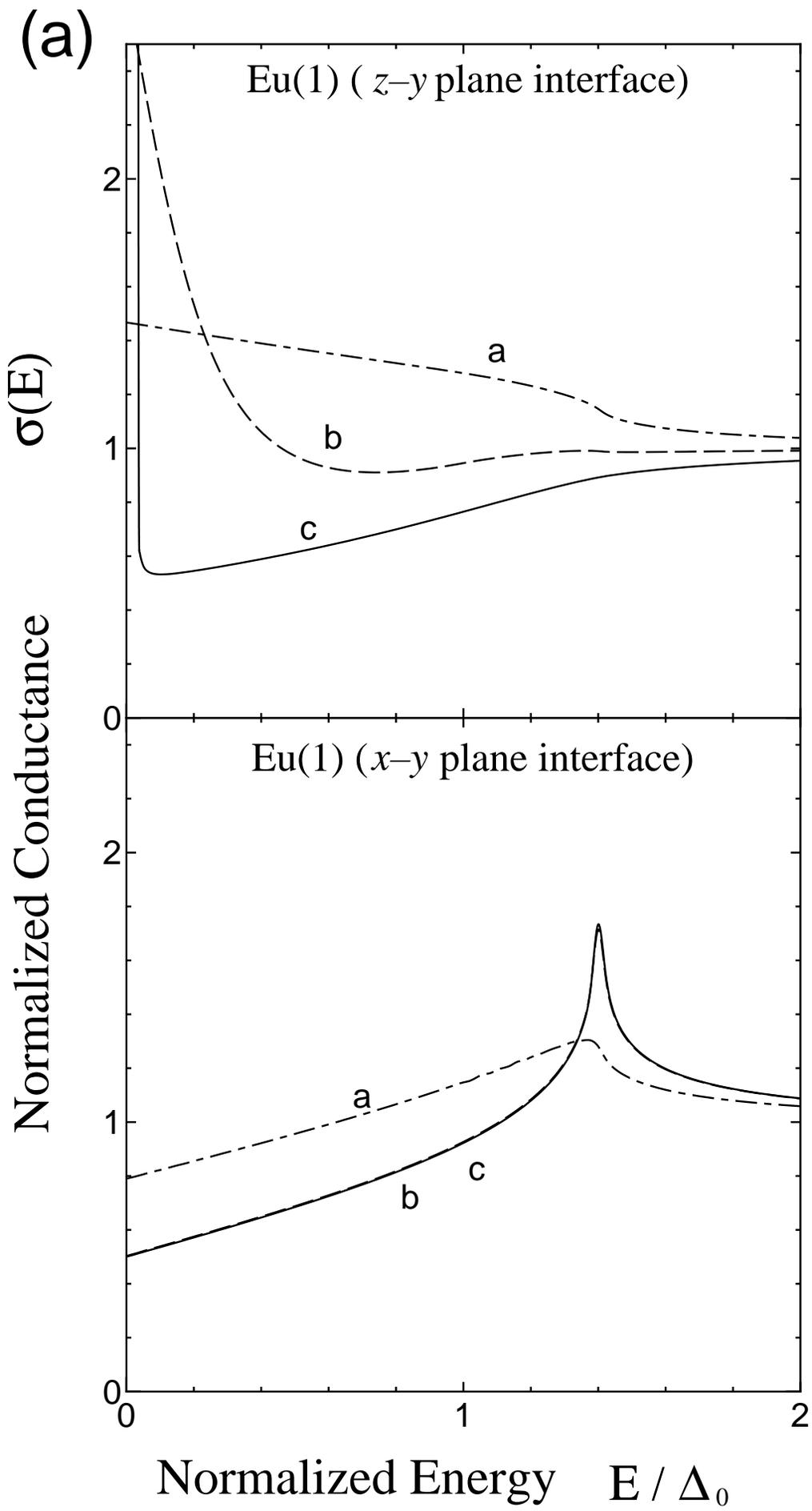
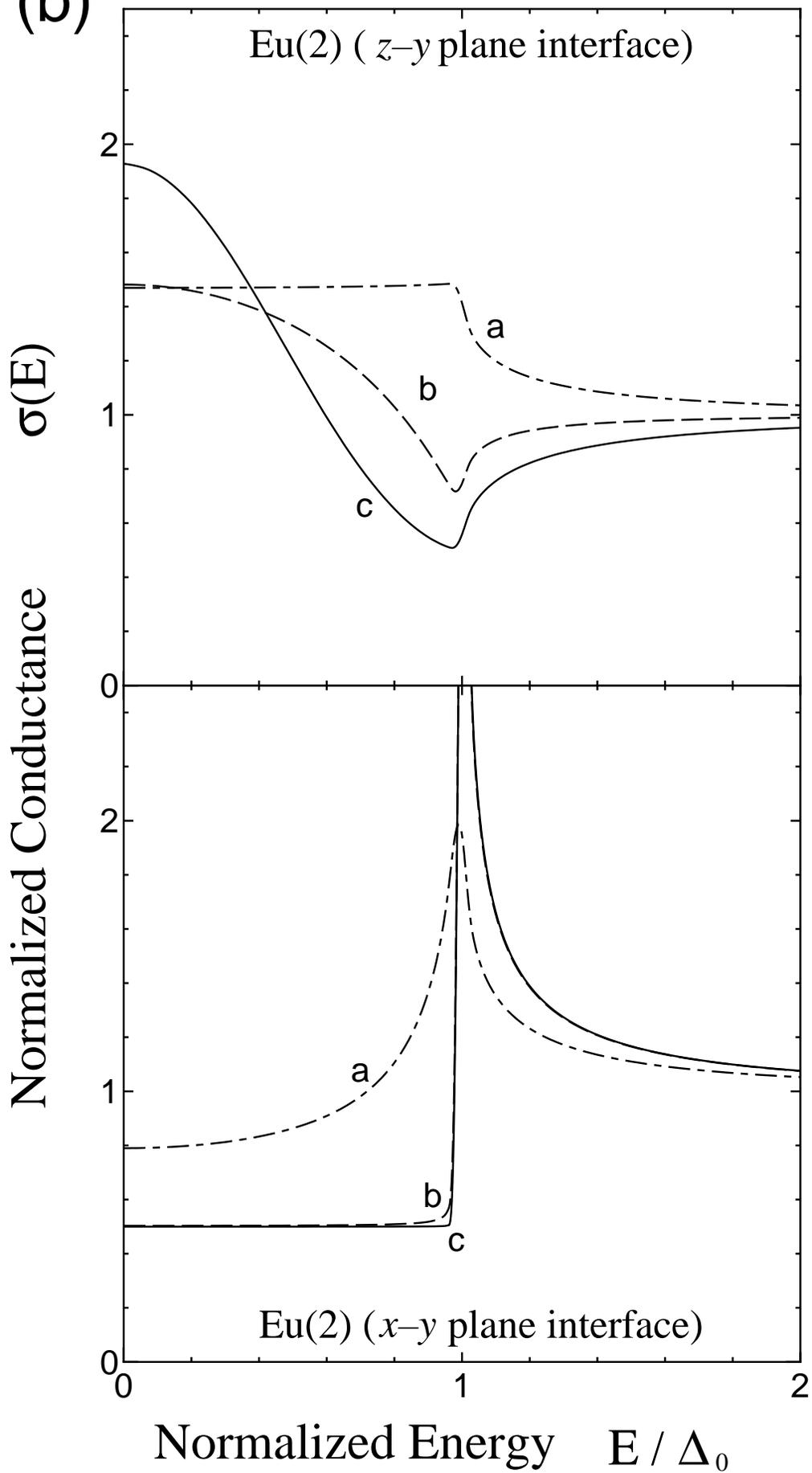
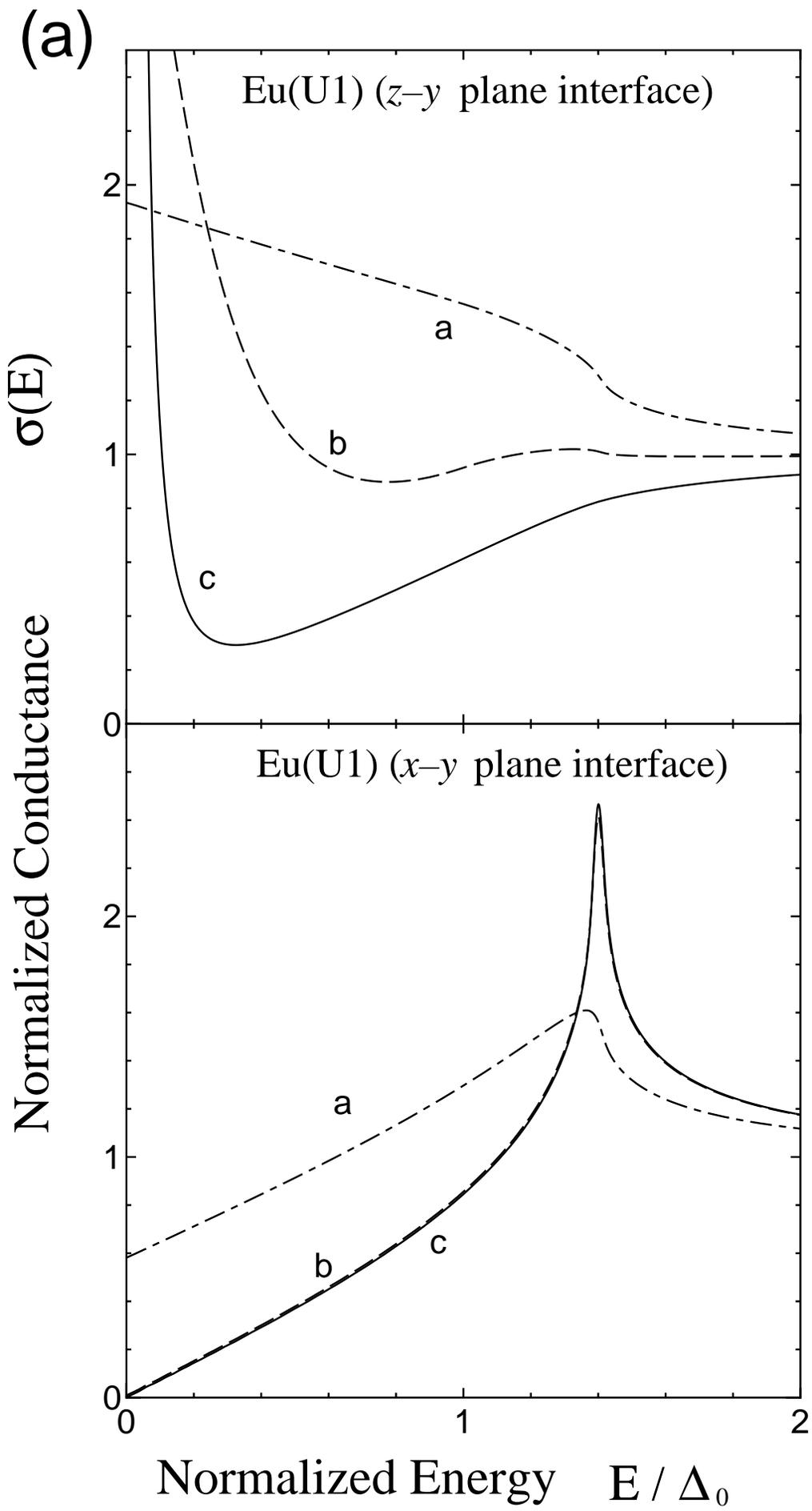
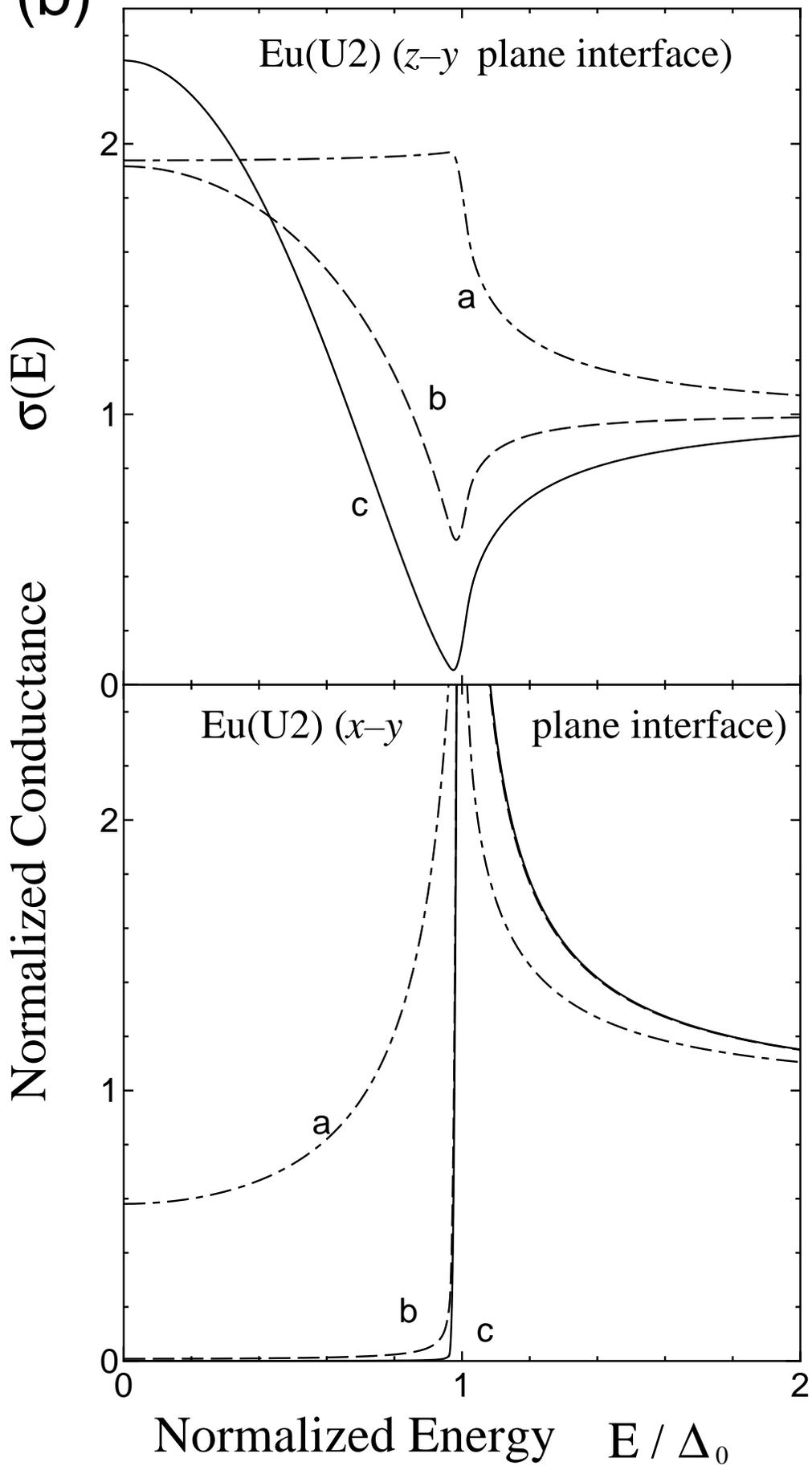
\end{document}